\documentclass[prl,twocolumn,superscriptaddress,showpacs,nofootinbib]{revtex4}
\usepackage{epsfig,graphicx}
\usepackage{flafter}
\usepackage{amsmath}
\usepackage{color}
\usepackage{bm}
\usepackage{amssymb}
\usepackage{epstopdf}
\usepackage{mathtools}
\usepackage{amstext}
\usepackage{amsfonts}
\usepackage{latexsym}
\usepackage{amsbsy}
\usepackage{makeidx}
\usepackage{latexsym}
\usepackage{mathtext}

\newcommand{\ket}[1]{\ensuremath{| #1 \rangle}}

\newcommand{\ex}[1]{\ensuremath{\langle #1 \rangle}}
\renewcommand{\vec}[1]{\boldsymbol{#1}}
\newcommand{\avg}[1]{\ensuremath{\langle #1 \rangle}}

\begin{document}

\title{A case study of spin-$1$ Heisenberg model in a triangular lattice}
\author{M. Moreno-Cardoner}
\affiliation{Departament de F\'isica, Universitat Aut\`{o}noma de Barcelona, E08193
Bellaterra, Spain}
\author{H. Perrin}
\affiliation{Departament de F\'isica, Universitat Aut\`{o}noma de Barcelona, E08193
Bellaterra, Spain}
\author{S. Paganelli}
\affiliation{International Institute of Physics, Universidade Federal do Rio Grande do Norte, 59012-970 Natal, Brazil}
\author{G. De Chiara}
\affiliation{Centre for Theoretical Atomic, Molecular and Optical Physics,
Queen's University Belfast, Belfast BT7 1NN, United Kingdom}
\author{A. Sanpera}
\affiliation{ICREA, Instituci\'o Catalana de Recerca i Estudis Avan\c{c}ats, E08011
Barcelona} 
\affiliation{Departament de F\'isica, Universitat Aut\`{o}noma de Barcelona, E08193
Bellaterra, Spain}

\date{\today}
\begin{abstract}
{We study the spin-$1$ model in a triangular lattice in presence of a uniaxial anisotropy field using a Cluster Mean-Field approach (CMF). The interplay between antiferromagnetic exchange, lattice geometry and anisotropy forces Gutzwiller mean-field approaches to fail in a certain region of the phase diagram. There, the CMF yields two supersolid (SS) phases compatible with those present in the spin-$1/2$ XXZ model onto which the spin-$1$ system maps. Between these two SS phases, the three-sublattice order is broken and the results of the CMF depend heavily on the geometry and size of the cluster. We discuss the possible presence of a spin liquid in this region.}
\end{abstract}
\maketitle

\section{I. INTRODUCTION}
The quest for realistic magnetic quantum models that support topological and spin liquid ground states is at the frontier of current research in condensed matter and in quantum information science. 
These exotic states of matter do not display local order (i.e. do not break any local symmetry), but possess long-range entanglement encoding a global structure that cannot be detected by any local measurement. This absence of local order makes them robust against local perturbations, and thus, promising candidates for technological applications \cite{Alet-Balents}.

The triangular lattice is the archetype for classical geometric frustration. This phenomenon is believed to play a crucial role in the stabilization of certain spin liquid quantum phases, such as the resonating valence bond states proposed by Anderson \cite{Anderson1987}. 
Such states are not only interesting because of exhibiting this exotic global order, but also they are expected to be related, when doped, to non-conventional superconductivity.

Here we analyze the spin-$1$ bilinear-biquadratic model in the triangular lattice with uniaxial field. Our study is at least two-fold motivated. On the one hand, the competition between dipolar and quadrupolar ordering involves a much richer physics than spin-$1/2$  systems, and leads to an effective spin-$1/2$ frustrated model for large values of the anisotropy field. 
On the other hand,  recent experimental results on some complex crystals where this model might be present show an anomalous behavior which could be related to a spin liquid state \cite{crystals}. Moreover,  quantum systems other than real crystals could 
simulate this model \cite{Imambekov}. For instance, ultracold atoms trapped in a triangular optical lattice \cite{lattices}. 

In contrast to the $S=1/2$ model in a triangular lattice, which has been thoroughly investigated \cite{{spin1/2a},{spin1/2b},{spin1/2c},{spin1/2d}}, conclusive results for the $S=1$ in this geometry are scarce, since powerful numerical techniques (e.g. DMRG, QMC) cannot be easily  applied here. First attempts to describe the full phase diagram of the model have been based on a mean-field (Gutzwiller) approach, which explicitly assumes a three-sublattice ordering  \cite{Toth2011}. While this approach properly sketches locally ordered phases, it cannot describe more complex states as those dominated by inter-site quantum correlations like spin liquids or supersolid phases.  In the absence of the anisotropy field, exact diagonalization (ED) (of up to 27 sites)  shows that three-sublattice order is always present, ruling out the existence of a spin liquid \cite{Lauchli2006}. However, for a finite value of the anisotropy field, controversial results regarding the presence of spin liquid phases have appeared using variational MC \cite{Bieri2012} and mean-field analysis in the triplon fermionic spin representation \cite{{Liu2010},{Serbyn2011},{Xu2012}}.

Motivated by these controversial results we investigate this model using a Cluster Mean-Field (CMF) approach. Within this approach, the spins in the lattice are grouped into small clusters which are exactly diagonalized, while each cluster is mean-field coupled to the other ones (see Fig.  \ref{Figure1}(a)) \cite{Yamamoto,Yamamoto2,Yamamoto3,Huerga}. 
This ansatz clearly constitutes a step forward as compared to the Gutzwiller mean-field approach since it exactly treats quantum correlations between all sites belonging to the same cluster. Undoubtedly, it also goes beyond the results obtained from the ED of an isolated plaquette of the same size. 

Before proceeding further with the details of our study, we briefly outline here our major findings: (i) the CMF provides corrections to the boundaries of the quantum phases obtained with the Gutzwiller approach, (ii) in stark contrast to the Gutzwiller approach, the CMF yields 
two distinct supersolid (SS) phases that for a large easy-axis uniaxial field map onto the supersolid phases present in the spin-$1/2$ XXZ model and (iii) between the two SS phases we find a region where our results strongly depend on the cluster geometry and size, as well as on the chosen initial cluster configuration. Moreover, as we will discuss in detail later, the presence of these instabilities is crucially linked to the underneath order imposed with the CMF approach. Such instabilities, on the other hand, could be compatible with the presence of a spin liquid, very much in the spirit of the instabilities shown by generalized spin wave theory in anisotropic triangular lattices in spin-$1/2$ systems \cite{Hauke}. Our CMF approach bounds quite precisely this region as compared to previous results \cite{Serbyn2011}. 

The paper is organized as follows. In Sec. II we introduce the model under study together with the CMF method, emphasizing its strength as compared to other approaches. In Sec. III we discuss the results of the spin-$1$ Heisenberg model, first reviewing the isotropic model  (III.A) and finally in presence of the single-ion anisotropy (III.B).  In Sec. IV we present our conclusions and open questions.

\section{II. THE MODEL  AND THE CLUSTER MEAN-FIELD METHOD}
The most general (n.n. interaction) rotationally invariant Hamiltonian for spin-$1$ particles reads: 
\begin{align}
H_{\mbox{\tiny{BB}}}(\theta)=\cos\theta \sum_{\ex{i,j}} \vec{S}_i \cdot \vec{S}_j + \sin \theta \sum_{\ex{i,j}} \left(\vec{S}_i \cdot \vec{S}_j\right)^2
\label{HBB}
\end{align}
where ${\vec{S}}=(S_x,S_y,S_z)$ are the usual spin $S=1$ operators and $\ex{i,j}$ runs over all possible pairs of first neighboring sites of the triangular lattice. The presence of a uniaxial field breaks the $SU(2)$ symmetry 
\begin{align}
H (\theta,D)=H_{\mbox{\tiny{BB}}}(\theta)+D \sum_i (S_i^z)^2.
\label{HBBD}
\end{align}
In the thermodynamic limit the system can display both, dipolar  and/or quadrupolar magnetic order, associated to the expectation values of 
$\mbox{\small{\avg{\vec{S}}}}$ and $\mbox{\small{\avg{\vec{Q}}}}=\mbox{\small{\avg {(S_x^2-S_y^2, (2{S}_z^2-S_x^2-S_y^2)/\sqrt{3}, \{S_x,S_y\}, \{S_y, S_z\}, \{S_z,S_x\})}}}$. Note that   
${\small{ (\vec{S}_i \cdot \vec{S}_j)^2= (\vec{Q}_i \cdot \vec{Q}_j - \vec{S}_i \cdot \vec{S}_j)/2+4/3}}$ and thus, the quadrupolar vector appears as a meaningful order parameter. The quadrupolar order is directly related to the anisotropy of spin fluctuations, which can be  characterized by the director $\hat{d}$, that defines the plane onto which the spin maximally fluctuates \cite{Lauchli2006}.\\

\noindent {\it{The CMF method: }} In the CMF method the lattice is divided into clusters of sites which are exactly diagonalized and coupled via a mean-field approach. The effective Hamiltonian acting on a cluster $c$ is then given by: 
\begin{equation}
{H}_{c}^{\text{eff}} = {H}_c+ \sum_{c'}{H}_{cc'}^{\text{mf}} \left(\ket{\psi_{c'}}\right)
\label{CMF}
\end{equation}
where ${H_c}$ is the (exact) Hamiltonian of cluster $c$ and ${H}_{cc'}^{\text{mf}}$ is the effective field  acting on cluster $c$ due to cluster $c'$:
\begin{equation} 
{H}_{cc'}^{\text{mf}} =\frac{1}{2}\sum_{\substack{\left<i,j\right>\\i\in c, j\in c'}}\left[(2\cos\theta - \sin\theta) \ex{ \vec{S}_j} \cdot \vec{S}_i
+ \sin \theta \ex{\vec{Q}_j} \cdot \vec{Q}_{i} \right] 
\end{equation}
Here the expectation values are evaluated over the self-consistent ground state $\ket{\psi_{c'}}$. We proceed as follows: starting with a random initial state for each cluster we iteratively solve (\ref{CMF}) until convergence is achieved. We use clusters of different geometries and sizes ($\sim$ 3-10 sites). 

\begin{figure}[t]
\centering
\includegraphics[width=8.5cm]{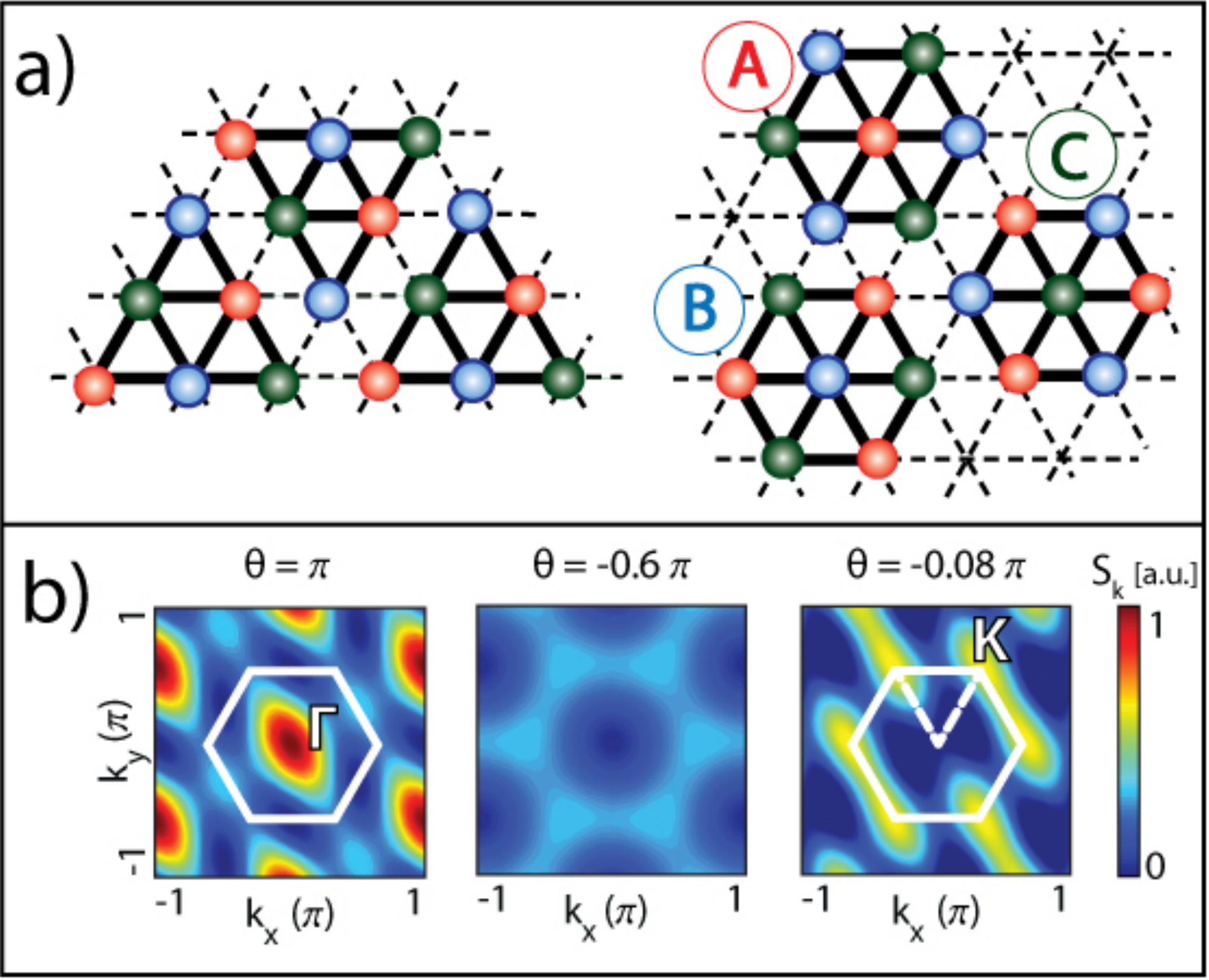}
\caption{(Color online). (a) The CMF method works by coupling via mean-field distinct clusters that are exactly diagonalized. Thick lines: quantum bonds within a cluster; dashed lines: mean-field bonds between clusters. Colors indicate the 3-sublattice order. For a 7 sites cluster the three configurations are not equivalent. (b) Dipolar structure factor ($S_S(k)$) as a function of reciprocal lattice vector $\vec{k}=(k_x,k_y)$ for $\theta=0$ (FM phase), $\theta=-0.6\pi$ (FQ phase) and $\theta=-0.08$ (AFM phase), calculated by ED for a 9 sites cluster. The $\Gamma$ and $K$ points in the first Brillouin zone are indicated. }
\label{Figure1}
\end{figure}

There is not a unique way of implementing CMF. Note that with this method, three-sublattice order (expected to naturally arise in an isotropic triangular lattice if the state is locally ordered) is not imposed on the final state. However, by considering different clusters in our ansatz, we give enough freedom to recover such ordering if present. For example, for a 6-sites cluster, a single configuration for all the plaquettes is sufficient, while for the 7-sites cluster we must use 3 different initial plaquettes to ensure that three-sublattice order can be recovered (see Fig. \ref{Figure1}(a)). In this respect, our approach differs from the one used in \cite{Yamamoto2,Yamamoto3}, where a sublattice background order is first imposed. After choosing the sublattice order, the external field to which the cluster is coupled is calculated by averaging over all sites that belong to the same sublattice. In addition, one has to average over all the non-equivalent ways of embedding a given cluster geometry in the sublattice structure (e.g. over the three cluster configurations sketched in Fig. \ref{Figure1}(a)). The results obtained with this second approach will be denoted by CMFav. 

A major effect of coupling the plaquettes with the others (which might act as an environment) is that certain symmetries (e.g. rotational invariance) are spontaneously broken, similarly to what happens in the thermodynamic limit. As a consequence, the CMF allows for a direct evaluation of the local order parameters $\mbox{\small{\avg{\vec{S}}}}$ or $\mbox{\small{\avg{\vec{Q}}}}$, in sharp contrast to the ED of a single cluster. For the latter, one evaluates instead the dipolar and quadrupolar structure factors which take into account spin-spin and quadrupolar-quadrupolar correlations respectively, as done in \cite{Lauchli2006} for the $D=0$ case. These are usually defined as: 
\begin{align}
S_S(\vec{k})&=\sum_{j}e^{i \vec{k}(\vec{r}_j-\vec{r}_0)} \avg{\vec{S}_0 \cdot \vec{S}_j} \\
S_Q(\vec{k})&=\sum_{j}e^{i \vec{k}(\vec{r}_j-\vec{r}_0)} \avg{\vec{Q}_0 \cdot \vec{Q}_j} 
\end{align}
where $\vec{r}_j$ runs over all sites of the cluster and $r_0$ is the central site. These functions reveal the order of a phase by showing  a maximum at certain value of the momentum $\vec{k}$. For example, for Ferro- (Antiferro-) magnetic order this maximum corresponds to the $\Gamma$ ($K$) point of the Brillouin zone, as illustrated  in Fig. \ref{Figure1}(b).

With the CMF method, a possible way to quantify quantum correlations in our system is through the renormalized linear entropy of a single spin with respect to the rest of the cluster, defined as:
\begin{equation} 
S^i_L= \frac{3}{2}\left(1-\text{Tr} \left[\rho_i^2\right]\right) = 1-\frac{3}{4} \left(\avg{\vec{S}_i}^2+\avg{\vec{Q}_i}^2\right)
\end{equation}
being $\rho_i$ the reduced density matrix of a single spin at site $i$ after tracing out the rest of the cluster. The average of this quantity over all cluster sites and configurations will be denoted by $S_L$. As it will be shown later, this parameter properly signals the phase transitions. 

On the other hand, the Gutzwiller ansatz for the triangular lattice is constructed by three independent fields ($\ket{\psi_A}$, $\ket{\psi_B}$ and $\ket{\psi_C}$) at each of the sublattices $\lambda$. Within this approach, and because any quantum correlation between sites is neglected, the local parameters always fulfill $\mbox{\small{\avg{\vec{S}}}}^2+\mbox{\small{\avg{\vec{Q}}}}^2=4/3$ ($S_L=0$), and therefore, phases are always locally ordered in one way or another. This ansatz fails to describe phases without three-sublattice order and/or dominated by quantum correlations. 

\section{III. RESULTS}
\subsection{A. ISOTROPIC MODEL ($D=0$)}
To test the performance of our CMF approach we start by revisiting the $SU(2)$ symmetric Hamiltonian (\ref{HBB}), whose ground state structure has been studied by means of the Gutzwiller ansatz \cite{Toth2011} as well as by ED \cite{Lauchli2006}. Within the CMF approach three-sublattice order is also found at any value of the parameter $\theta$. In particular, four phases with local magnetic order emerge: {\it{Ferro-Magnetic  order}} (FM) for $\theta\in(\pi/2,5\pi/4)$, {\it{Ferro-Quadrupolar  order}} (FQ) for $\theta\in (5\pi/4, \Theta)$, {\it{Antiferro-Magnetic order}} (AFM) for $\theta \in(\Theta, \pi/4)$, and {\it{Antiferro-Quadrupolar order}} (AFQ) for $\theta\in(\pi/4,\pi/2)$. The quadrupolar phases do not show net magnetization at each site but quadrupolar order. In the Ferro (F) phases the vectors $\vec{S}$ ($\vec{Q}$) associated to dipolar (quadrupolar) magnetic order are aligned, whereas for the Antiferro (AF) phases they are all contained in the same plane and form $120^\circ$. 

\begin{figure}[h!]
\centering
\includegraphics[width=8cm,clip=]{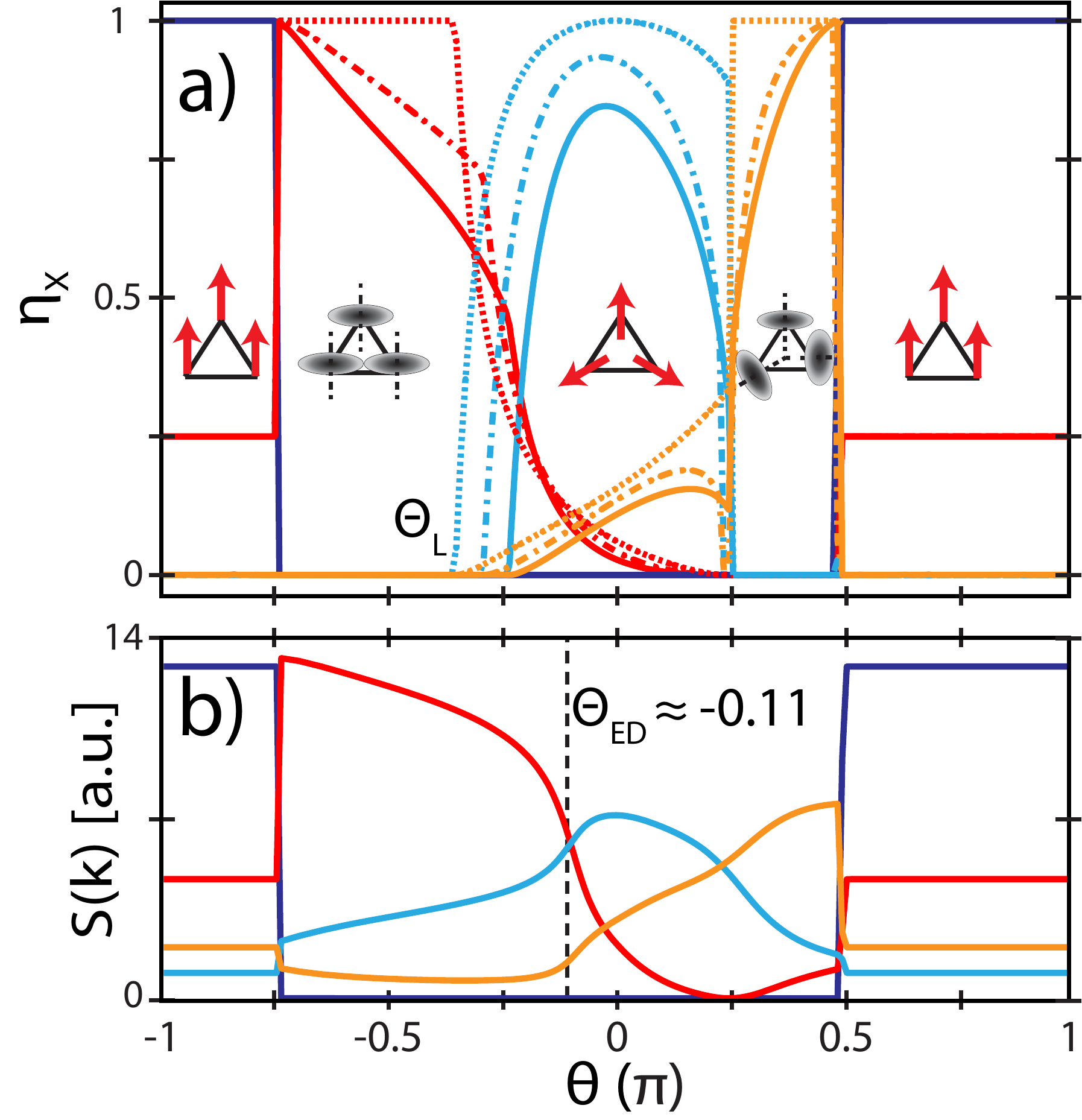}
\caption{(Color online). $D=0$ results. (a) Local order parameter $\eta_{\text{FM}}$ (blue), $\eta_{\text{AFM}}$ (cyan), $\eta_{\text{FQ}}$ (red) and $\eta_{\text{AFQ}}$ (orange) (see text), calculated with the CMF method for clusters of 3 and 9 sites (dotdashed and solid respectively). The dotted line is the Gutzwiller ansatz for comparison. The FQ-AFM transition point ($\Theta_L$) strongly renormalizes with the cluster size.  (b) Structure factors $S_S(\vec{k}=\Gamma)$ (blue), $S_S(\vec{k}=K)$ (cyan), $S_Q(\vec{k}=\Gamma)$ (red) and $S_Q(\vec{k}=K)$ (orange) calculated by ED for $L=12$. The FQ-AFM transition is at $\Theta_{\text{ED}}  \approx -0.11 \pi$ (see also \cite{Lauchli2006}).}
\label{Figure2}
\end{figure}

A convenient way to characterize these three-sublattice locally ordered phases is by the following order parameter:  
\begin{equation}
\eta_X=\sum_{j=1}^3 \sum_{i\in \lambda_j} \mathcal{R}(\phi_j^X) \vec{A}_i 
\end{equation}
where $X$ denotes the phase, $i$ runs over all lattice sites of the cluster and $\lambda_j$ (with $j=1,2,3$) denote the three sublattices. $\mathcal{R}(\phi_j^X)$ is a rotation in real space of the vector $\vec{A}_i$, with an angle that depends on the sublattice $j$ that contains site $i$. For the F and AFM phases $\vec{A}=\avg{\vec{S}}$ and $\phi^\text{F}_j=0$, $\phi^{\text{AF}}_j=0,\pm 120^\circ$, whereas for the FQ and AFQ phases  $\vec{A}=\avg{\vec{Q}}$ and $\phi_j^{\text{FQ}}=0$ and $\phi_j^{\text{AFQ}}=0,\pm 120^\circ$. This rotation is performed on the plane containing the spins (or quadrupolar vectors). These order parameters are shown in Fig. \ref{Figure2}(a)  for several cluster sizes.

The Gutzwiller ansatz already yields the correct phases and for all but the FQ-AFM transition the phase boundaries coincide with those obtained by ED. For the latter (for clusters of 9 sites and larger) this yields $\Theta_{\text{ED}} \approx -0.11\pi$  (see Fig. \ref{Figure2}(b) and also  \cite{Lauchli2006}), whereas with the Gutzwiller ansatz the transition occurs at $\Theta_{\text{Gutz}} = -0.35\pi$. The inclusion of correlations in the CMF approach yields a critical point that shifts to smaller values in modulus as the cluster size increases.  With our available cluster sizes, we perform a linear fit as in \cite{Yamamoto2,Yamamoto3} on the value $\Theta_L$ as a function of the parameter $x=N_b/(L \times z/2)$, where $N_b$ and $z$ are the number of quantum-mechanical bonds in the cluster and coordinate number  respectively ($x=0$ for Gutzwiller and $x=1$ in the thermodynamic limit). This yields the transition at $\theta \approx -0.17\pi$ in the thermodynamic limit, as shown in Fig. \ref{Figure4}(a). 

\subsection{B. MODEL WITH SINGLE-ION ANISOTROPY ($D\neq 0$)}
The presence of a uniaxial anisotropy field along an arbitrary axis, e.g. $\hat{z}$-axis, drastically changes the nature of the ground state. When the anisotropy does not dominate over the exchange coupling the system is far from being well understood.  Let us first examine the already known limiting regimes.\\

\begin{figure}[t]
\centering
\includegraphics[width=9cm]{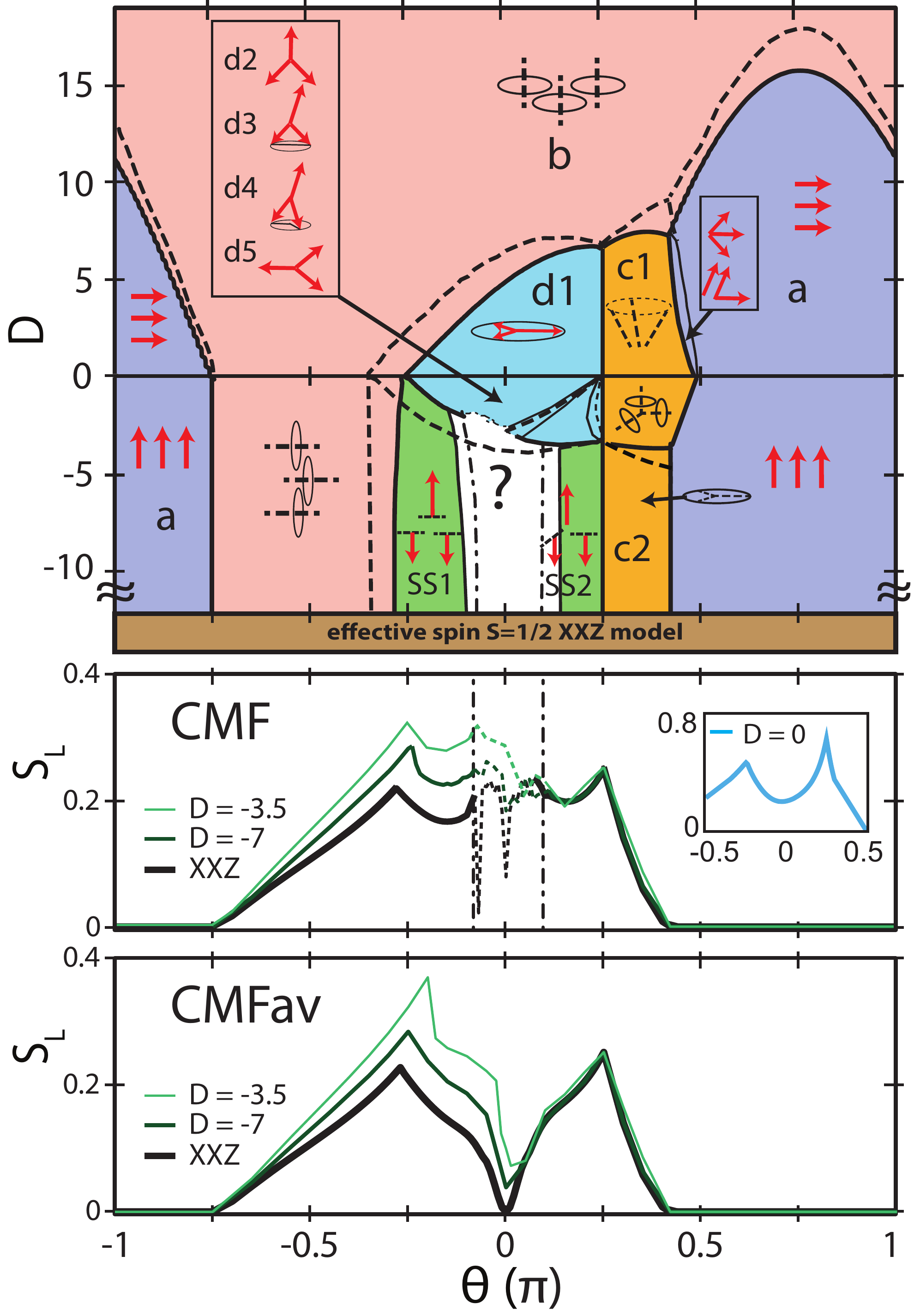}
\caption{(Color online) Top: CMF phase diagram using a triangular cluster of 6 sites (solid).  Arrows and broken lines denote $\vec{S}$ and $\hat{d}$ directions respectively. Gutzwiller mean-field boundaries are also shown for comparison (dashed lines). For $D<0$, two supersolid phases (SS1 and SS2) not present in the mean-field solution arise. In between, there is a region where CMF is unstable and results strongly depend on the imposed sublattice order in the CMF (see text). The dotdashed line denotes this boundary for a 10 sites cluster. Bottom: Quantum correlations with CMF for $L=10$  measured by linear entropy $S_L$ for different values of negative $D$ (decreasing value from thin to thick, thicker line is spin-$1/2$ XXZ model). If no order is imposed (middle panel) the method shows an  instability between the two SS phases, whereas by imposing three-sublattice order (CMFav) continuity is recovered (bottom panel). The inset shows the result for $D=0$, where CMF is always stable.}
\label{Figure3}
\end{figure}

\noindent \textit{Limit $D\gg1$}. For very large and positive uniaxial field the ground state is the product state $\ket{0,. . . ,0}$, which has only quadrupolar order (FQ) at each site (large-D phase). The corresponding vector directors $\hat{d}$ are oriented along $\hat{z}$ and spin fluctuations restricted to the normal plane to $\hat{z}$ as depicted in Fig. \ref{Figure3}. \\

\noindent \textit{Limit $D\ll-1$}. In the opposite limit, the local $\ket{0}$ component of the spin-$1$ $( \ket{S_i = 1, S_{zi}=-1,0,1} )$ is adiabatically suppressed and the two left components per site can be regarded as a pseudospin-$1/2$, when mapping  $\left\lbrace \ket{1},\ket{-1}\right\rbrace \longleftrightarrow \left\lbrace \ket{\uparrow},\ket{\downarrow} \right\rbrace$. Strictly in the limit $D=-\infty$, any configuration in this subspace has the same energy and there is a macroscopic degeneracy. For large but finite values of $|D|$ the degeneracy is broken by the spin exchange terms. Taking into account that $ \mathcal{P} (\vec{S}_i \cdot \vec{S}_j)^2 \mathcal{P} =2\left(\sigma_i^{x} \sigma_j^{x}+\sigma_i^{y}\sigma_j^{y}\right)$ and $\mathcal{P} \vec{S}_i\cdot \vec{S}_j \mathcal{P} = \vec{\sigma}_i \cdot \vec{\sigma}_j$  (where $\mathcal{P}$ is a projector to the subspace with no local component $\ket{0}$), the effective Hamiltonian in this limit corresponds to the spin-$1/2$ XXZ model :  
\begin{equation}
{H}_{\text{XXZ}}=\sum_{<i,j>}\left[ J_\perp \left(\sigma^x_i \sigma^x_j + \sigma^y_i \sigma^y_j\right) + J_z \sigma^z_i \sigma^z_j\right]
\label{Eq:XXZmodel}
\end{equation}
with parameters depending on $\theta$ as  $2J_\perp = \sin \theta$ and $2J_z = (2\cos\theta-\sin \theta)$ \cite{DeChiara_spin1}. 
It is easy to see that magnetization along the $\hat{z}$-direction in the $S=1/2$ model directly maps into dipolar order along the same axis for $S=1$, whereas the transverse magnetization maps into quadrupolar order for the latter. Specifically, the magnetization along $\hat{x}$ and $\hat{y}$-axis in the spin-$1/2$ case map respectively to the components $Q_1$ and $Q_3$ of the quadrupolar vector.
The spin-$1/2$ model on the triangular lattice has been addressed by several approaches  (like QMC or variational methods)  \cite{{spin1/2a},{spin1/2b},{spin1/2c},{spin1/2d}} and predicts the presence of two distinct  supersolid phases (which we denote by SS1 and SS2). These two phases are characterized by long range $\sqrt{3} \times \sqrt{3}$ crystal order coexisting with superfluidity, which in spin language dubs to long range $\sqrt{3} \times \sqrt{3}$ order in the spin $z$-component and non-vanishing magnetization in the perpendicular plane. This transverse  magnetization differs in the two phases: for SS1 it takes the same value at each sublattice  $(m_\perp,m_\perp,m_\perp)$, whereas for SS2 it corresponds to a configuration $(0,m_\perp,-m_\perp)$.\\

\noindent\textit{Intermediate values of $D$.} For $D>0$, the phase diagram can be qualitatively well characterized by the Gutzwiller approach as shown in Fig. \ref{Figure3}, where phase boundaries from Gutzwiller ansatz (dashed lines) and CMF approach (solid) are drawn. 
Such characterization notably fails in a region of $D <0$. The phases depicted in Fig. \ref{Figure3} can be summarized as follows:\\

\noindent\textit{Region a: Ferromagnetic (FM) order.}--- The FM phase persists for any  $D \leq 0$, showing magnetization along the field direction, and a double degeneracy due to $Z_2$ symmetry. When crossing to the $D>0$ side, the spins point instead along any perpendicular direction to the field. As $D$ increases the magnetization continuously decreases until it vanishes and only FQ order remains. 
Near $\theta\approx 0.5 \pi$,  there is a small region presenting ``pathological''  transverse magnetization.\\

\noindent\textit{Region b: Ferroquadrupolar (FQ) order.}--- This phase shows only pure quadrupolar order with the corresponding directors at different sites being parallel. The directors are aligned with (perpendicular  to) the field axis for $D>0$ ($D<0$), while they can take any orientation when crossing  the $SU(2)$ symmetry point $D=0$. The FQ order becomes more stable when including quantum correlations as it can be seen in Fig. \ref{Figure2} by comparing the boundaries of the mean-field with those of the CMF.\\

\noindent \textit{Regions c: Antiferroquadrupolar (AFQ) order.}--- As in region {\it{b}}, there is no dipolar but quadrupolar local order. Now the vector directors have different orientation at each of the three sublattices, and their angle continuously varies with $D$. In particular, for $D<D_{c1}$, the directors are no longer parallel and start to form an umbrella configuration (region {\it{c}}1 in Fig. \ref{Figure2}), until they are mutually orthogonal for $D=0$. By further decreasing $D$ the angle still increases until they all lie within the $\hat{x}\hat{y}$-plane  at $D=D_{c2}$. For $D<D_{c2}$ (region {\it{c}}2) they remain in this configuration, subtending a $120^\circ$ angle. Both critical fields ($D_{c1}$ and $D_{c2}$) depend on $\theta$.\\

\noindent\textit {Regions d: Antiferromagnetic (AFM) order.}--- This region sustains different quantum phases. They are characterized by local dipolar order which is not ferromagnetic, but still show a three-sublattice structure, and a non-vanishing magnetization $\avg{\vec{S}^\perp}\neq 0$ at least at one of the three sublattices. For $D>0$ (region {\it{d}}1), the ground state is in a $120^\circ$-N{\'e}el ordered phase (similarly to the $D=0$ case, but now with the spins contained in the $\hat{x}\hat{y}$-plane). For $D<0$, there are at least four different phases that are already observed within the mean-field approach. They are found consecutively when increasing $\theta$ from negative to positive values at fixed $D$ and characterized through the net magnetization along the parallel and perpendicular directions to the field axis ($\avg{\vec{S^{z}_{\text{tot}}}}$ and $\avg{\vec{{S}^{\perp}_{\text{tot}}}}$ respectively)  as follows: {\it{d}}2) $\avg{\vec{{S}^{z}_{\text{tot}}}}\neq 0$, $\avg{\vec{{S}^{\perp}_{\text{tot}}}}=0$; {\it{d}}3) $\avg{\vec{{S}^{z}_{\text{tot}}}}\neq 0$, $\avg{\vec{{S}^{\perp}_{\text{tot}}}} \neq0$; {\it{d}}4) $\avg{\vec{{S}^{z}_{\text{tot}}}}= 0$, $\avg{\vec{{S}^{\perp}_{\text{tot}}}} \neq 0$; {\it{d}}5) the spins representing each sublattice are not contained in the same plane, leading to a non-zero value of the spin-chirality scalar defined as $\kappa=\avg{\vec{S}_1\cdot (\vec{S}_2\times \vec{S}_3 )}$.\\

\noindent\textit {Regions SS1 and SS2: Supersolid Phases.}--- By decreasing $D$ ($D\lesssim -3.5$) in region {\it{d}}, a new order characterized by $\avg{\vec{S}^\perp}=0$ at any lattice site emerges. In this region the Gutzwiller ansatz fails, because the Hamiltonian is reduced to the classical XXZ model, which exhibits phases with qualitatively different local order and even non-trivial macroscopic degeneracies that are broken in the quantum model. Indeed, already at a finite value of the easy-axis anisotropy field the local $\ket{0}$ component is completely suppressed, and thus, the model reduces to the spin-$1/2$ XXZ model, exactly as in the $D\rightarrow -\infty$ limit. Furthermore, within the mean-field ansatz this model is equivalent to the classical one. In contrast, the CMF yields two phases that possess magnetic and quadrupolar order coexisting and which continuously connect with the mapped supersolid phases (SS1 and SS2) expected for the effective spin-$1/2$ XXZ model in the large easy-axis anisotropy limit ($D\ll -1$) \cite{{spin1/2a},{spin1/2b},{spin1/2c},{spin1/2d}}. Notice that in the triangular lattice the two SS-phases cannot be mapped onto each other ($|J_\perp|\nleftrightarrow -|J_\perp|$). Thus, it is not surprising that the prolongation of these phases to finite values of $D$ is different.

In stark contrast to the Gutzwiller case, quantum correlations prevent now the local $\ket{0}$ component to be totally suppressed in this region. In particular, we find that for $\theta >0 $, the $S=1$ model approaches to the equivalent spin $S=1/2$ model much faster than for $\theta<0$ in the SS1 phase. Hence, the suppression of the local $\ket{0}$ component is much smoother in the SS1 than in the SS2 phase. This can be appreciated in the behavior of the linear entropy $S_L$ shown in Fig. \ref{Figure3} (bottom). 

\begin{figure}[h!]
\centering
\includegraphics[width=8.75cm]{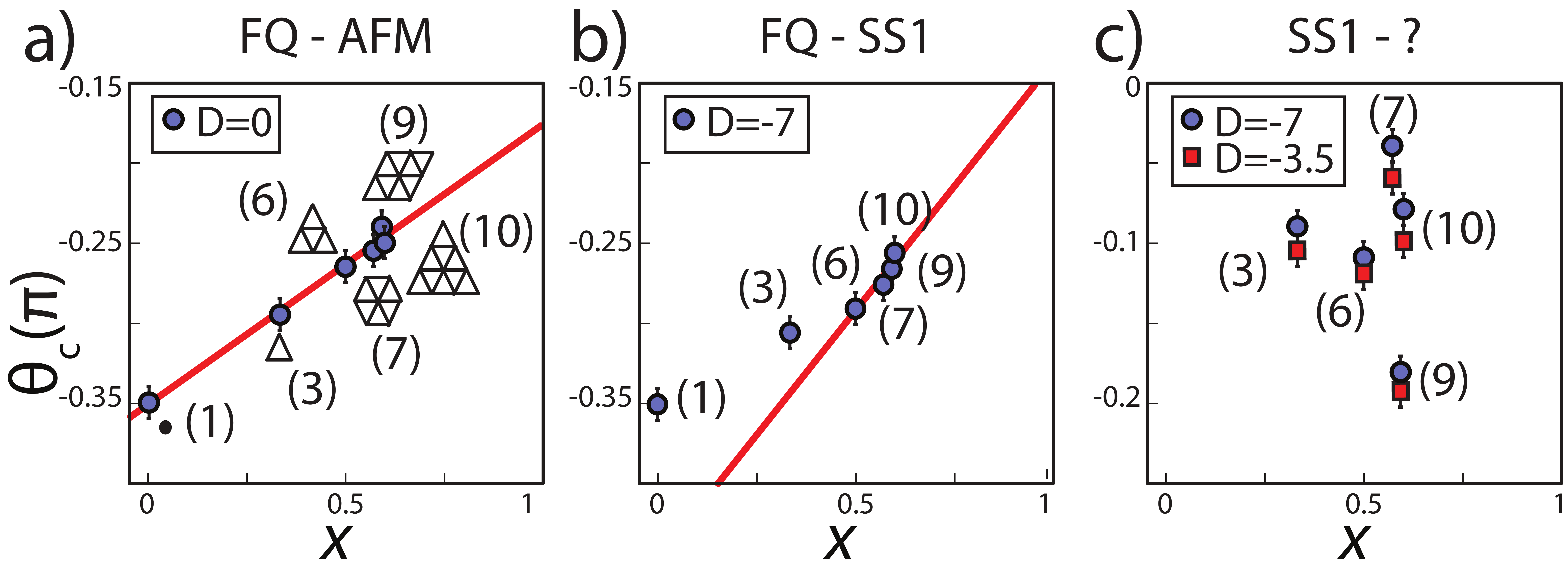}
\caption{Scaling of different phase boundaries for different values of $x$  (see text) within CMF:  (a) The $FQ-AFM$ boundary at $D=0$, (first point corresponds to the Gutzwiller value for comparison),  (b) $FQ-SS1$ phase at $D=-7$ (first point corresponds to the Gutzwiller value for comparison), (c) boundary between SS1-? phase at $D=-3.5$ (squares) and $D=-7$ (circles) showing lack of scaling.} 
\label{Figure4}
\end{figure}

With our method we find a region between the two supersolid phases (indicated by the symbol ?) where three-sublattice order is,  in general, broken. There, the results of our CMF show a strong dependence on both the initial state for a given cluster configuration and on the geometry and size of the chosen clusters. The instability is in fact linked to an enhancement of quantum correlations when approaching this region. As shown in Fig. \ref{Figure3} (middle panel), after a sudden increase, $S_L$ strongly fluctuates in this region. We emphasize that only in this part of the phase diagram the method shows such behavior. Moreover, the dependence on the cluster shape can be appreciated by the absence of a clear scaling behavior of the boundary separating SS1-?.  In Fig. \ref{Figure4} we depict this scaling as a function of the connecting parameter $x$ defined in Sec. III.A. While transitions FQ-AMF for $D=0$ and  FQ-SS1 at $D=-7$ accept a linear fit in $x$, the transition SS1-? clearly does not support such a fit. 

It is tempting to regard such instability in this region of the phase diagram as a signature of disorder in the ground state. In order to check such possibility, we study with our method the equivalent XXZ spin-$1/2$ model which governs the physics of our system at very large easy-axis anisotropy. For such model there exists numerical evidence that no spin liquid is stable and that the two supersolid phases are continuously connected \cite{{spin1/2a},{spin1/2b},{spin1/2c}}. In this case, a similar instability is found, as shown in Fig. \ref{Figure3} (middle panel). However, if instead we apply a CMF approach with imposed three-sublattice order in the external mean-fields as explained in Sec. II (CMFav), the instability is in both cases (spin-$1$ and spin-$1/2$) removed, and the function $S_L$ becomes smooth. The results using the CMFav are shown in \ref{Figure3} (bottom panel).

For completeness, we have applied the same method in the anisotropic triangular spin-$1/2$ XY and Heisenberg models, for which it is known that the ground state is a spin-liquid for some values of the coupling anisotropy \cite{Schmied,Hauke,Reuther,Yunoki,Hayashi,Heidarian}. Interestingly enough, for those regions we have also found that the CMF becomes unstable, but the instability is again removed by imposing the appropriate underlying sublattice order in the CMFav.\\

\section{IV. CONCLUSIONS}
We have obtained the (CMF) phase diagram for spin-$1$ Heisenberg model with uniaxial anisotropy in the triangular lattice. Our approach goes well beyond Gutzwiller mean-field and predicts two supersolid phases compatible with the effective spin-$1/2$ model expected in the limiting regime of large easy-axis anisotropy. Between them there is a phase whose nature remains elusive. It is signaled by a strong discrepancy between the results obtained with different cluster configurations and initial conditions. 

Therefore, the present calculation does not allow us to rule out (neither to prove) the existence of a spin liquid in this region of the phase diagram, but permits to bound quite precisely where quantum correlations become crucial for phase stability. It remains an open question why in this region of the phase diagram convergence is only achieved if three-sublattice order is imposed, while this stringent condition is not required for all other ordered phases present in the model. This issue could be addressed with the CMF  by using much larger clusters. 

\begin{acknowledgements}
We acknowledge financial support from: MINECO FIS2008-01236 (Spain), Grant ID SGR2009-1289 (Catalonia), Grant ID 43467 (Templeton Foundation), Grant ID EP/L005026/1 (EPSRC),
Grant ID EP/K029371/1, and Grant ID 618074 (EU Project TherMiQ). S.P. acknowledges partial support from MCTI and UFRN/MEC (Brazil). Fruitful discussions with A. L\"{a}uchli, M. Lewenstein and C. Lhuillier are appreciated.
\end{acknowledgements}

\end{document}